\documentclass[useAMS,referee,doublespacing,usenatbib,usegraphicx]{mn2e}
\usepackage{times}

\title[Modified surface redshift of pulsars and magnetars]
{Modified surface redshift of pulsars and magnetars by magnetized plasmas and vacuum polarization}
\author[Yuee Luo and Peiyong Ji]
{Yuee Luo$^{1}$, Zhigang Bu$^{1}$, Wenbo Chen$^{1}$, Hehe Li$^{1}$ and Peiyong Ji$^{1,2}$\thanks{E-mail:pyji@staff.shu.edu.cn.}\\
$^{1}$Department of Physics, Shanghai University, Shanghai 200444, China\\
$^{2}$The Shanghai Key Laboratory of Astrophysics, Shanghai 200234, China}

\begin{document}

\pagerange{\pageref{firstpage}--\pageref{lastpage}}
\maketitle
\label{firstpage}

\begin{abstract}
The propagation of emissions in the relativistic streaming pair plasmas of pulsars and magnetars
is studied within the framework of Gordon effective metric theory, and the effect of vacuum polarization
on the radiation spectra is examined particularly.
It is found that the electromagnetic and kinetic effects of streaming magnetized plasmas and the
vacuum polarization effect can shift the radiation spectra of stars. The modification of redshifts
of spectra can reach the same magnitude as the gravitational redshift in particular cases.
Moreover, the redshifts induced by the media are anisotropic and associated with wave modes.
For the O-mode, the modification of redshifts is dependent on the frequency of the radiation.
For lower-frequency radiation, the modification is dominated by the plasma effect,
while for higher-frequency radiations, by the vacuum polarization effect.
For the X-mode, the modification is non-dispersive and dominated by the vacuum polarization effect.
The vacuum polarization also has a significant effect on the redshifts of X-ray emissions in magnetars.
\end{abstract}


\section{Introduction}
Since the discovery of pulsars by Hewish in 1927, people established different models to describe
the dynamics of pulsars, such as the neutron star and strange star models \citep{GO69,MA01,ZH00,XU01,US01}.
When solving the structural model of stars by using the equation of state and equation of hydrostatic equilibrium,
one often needs to calculate the mass-to-radius ratio ($M/R$) of stars.
Theoretically, the mass-to-radius ratios are different for the neutron star and strange star.
The former satisfies $M\propto R^{-3}$, while the latter is $M\propto R^{3}$.
Thus, the stellar mass-to-radius ratio ($M/R$) can manifest the constituents of the core of stars.
According to present researches, the most straightforward method of determining
the mass-to-radius ratio is by measuring the gravitational redshifts $z$ of spectral lines,
which gives $M/R=(c^2/2G)[1-(1+z)^{-2}]$. \citet{CO02} reported the discovery of significant
absorption lines in the spectra of 28 bursts of the low-mass X-ray binary EXO 0748-676, with most
of the features associated with $\textmd{Fe}^{26}$ and $\textmd{Fe}^{25}$ $n=2-3$ and
$\textmd{O}^{8}$ $n=1-2$ transitions, all at a redshift $z=0.35$. Identification of absorption
lines was also achieved by \citet{SA02} using Chandra observations of the isolated neutron
star 1E 1207.4-5209. The lines were found to correspond to energies 0.7 and 1.4 keV, which are
interpreted as the atomic transition of singly ionized helium in a strong magnetic field.
The inferred redshift is $0.12-0.23$. \citet{TI05} discovered a redshifted X-ray emission
line centered at about 19.2 ${\AA}$ in the symbiotic neutron star binary 4U 1700+24.  They gave two
possible interpretations for this line: the Ly-$\alpha$ transition of $\textmd{O}^8$ at redshift $z\simeq0.012$ or
the $\textmd{Ne}^9$ triplet at redshifts $z\sim0.4$.
These observations provide an opportunity to measure the mass-to-radius ratio and constrain the
equation of state of the superdense matter.

However, we should notice that the redshifts of spectral lines are not only attributed to the gravitational field,
but also to other effects. A very interesting example is provided by the discovery by Ibrahim et al.
\citep{IB02,IB03}. They reported the discovery of cyclotron resonance features in the source SGR 1806-20,
said to be a magnetar candidate by \citet{KO98}. The features consist of a narrow 5.0 keV absorption
line with modulation near its second and third harmonics.
When these features are interpreted as electron-cyclotron line originating close to the surface of a typical
$B\approx10^{12}-10^{13}\textmd{G}$ neutron star in the context of accretion models,
it wound require a very large gravitational redshift $z>1.5$, with a mass-to-radius ration
$M/R>0.3M_\odot$ $\textmd{km}^{-1}$. Such values are inconsistent with neutron stars,
where $M_\odot$ is the mass of the Sun.
Although, the 5.0 keV feature is found to be a proton-cyclotron fundamental in the magnetar picture,
there still exists a possibility that the redshifts of spectral lines are not pure gravitational redshifts,
so other effects should be taken into consideration in interpreting spectral features.
\citet{MO04,MOS04} pointed out that non-linear electrodynamics effects
produced by the superstrong magnetic field of pulsars can modify gravitational redshifts.
\citet{JI08} also suggested that the magnetized plasma can affect redshifts of spectral lines.
Hence, a correct procedure to estimate the mass-to-radius ratio and the equation of state of a
compact star from the gravitational redshifts demands a separation of non-gravitational redshifts
from the pure gravitational ones.

In this paper, we suggest that the gravitational redshifts of spectral lines are modified when
the radiations pass through the medium around pulsars and magnetars.
It is well known that pulsars and magnetars possess extremely strong magnetic fields and are
surrounded by relativistic, magnetized electron-positron pair plasmas. Furthermore, the vacuum
becomes birefringent in the strong magnetic field \citep{HE36,AL71}. The redshifts of spectral lines
due to the plasma effect and vacuum polarization are studied together using Gordon effective metric theory \citep{GO23,EH67}
that is generalized to weak-dispersive media. In this method, the gravitational effect and
the electromagnetic and kinetic effects of the medium are together geometrized in terms of an effective metric.
The results show that the plasma and vacuum polarization effects play a key role in the redshifts of
spectral lines, and the effects are magnified by the kinetic effect of relativistic streaming plasmas.
Unlike the gravitational redshift being isotropy, the redshifts due to the media are
different for two wave modes: the ordinary mode (O-mode) and the extraordinary mode (X-mode).
Such property can help us to discriminate the redshifts produced either by the gravitational field
or by the effect of the media.

The paper is organized as follows. In Section 2, the redshifts of spectral lines produced at the surface of
pulsars and magnetars are analyzed using the Gordon metric theory.
In Section 3, the modification of the gravitational redshifts, caused by magnetized
pair plasmas and vacuum polarization, is calculated in detail.
Finally, some discussions and conclusions are given in Section 4.

\section{modification of the gravitational redshift}
The change in the frequency of electromagnetic radiation in a gravitational field is
predicted by Einstein's general relativity. The shifting of photon frequency is dependent
on the distribution of localized gravitational field, and the redshift $z$ measured
by an observer is defined in terms of frequencies
\begin{equation}
1+z=\frac{\nu_e}{\nu_o},
\label{1}
\end{equation}
where $\nu_e$ and $\nu_o$ are photon frequencies at the emitter and observer respectively.
Particles moving in the space-time with static metrics are governed
by the localized energy conservation law $|g_{00}(r)|^{1/2}E_\rmn{local}=\textmd{const}.$ \citep{MI73},
where $E_\rmn{local}$ is the localized energy of particles and $g_{00}(r)$ denotes the time-time
component of metric. For photons, $|g_{00}(r)|^{1/2}\nu_\rmn{local}=\textmd{const}$.,
so Eq. (\ref{1}) can be written as
\begin{equation}
1+z=\sqrt{\frac{g_{00}(o)}{g_{00}(e)}},
\label{2}
\end{equation}
where $g_{00}(e)$ and $g_{00}(o)$ are the time-time components of the gravitational metrics
at the emitter and the observer respectively. In the spherically symmetric Schwarzschild
gravitational field, $g_{00}(e)=1-2GM/(c^2R)$,
where $G$, $M$ and $R$ are the gravitational constant, the mass and the radius of stars respectively.
Thus, the gravitational redshift of the spectral line emitted from the stellar surface can be expressed as
\begin{equation}
1+z=\left[1-\frac{2GM}{c^2R}\right]^{-\frac{1}{2}}.
\label{3}
\end{equation}
For a typical pulsar with the mass $M\approx1.4M_\odot$ and the radius $R\sim 10$ $\textmd{km}$,
the gravitational redshift is $z\approx0.31$.

The gravitational redshift of spectral lines of the emission is not the only effect.
The electromagnetic and kinetic effects of the plasmas and the effect of vacuum polarization
on the emission should also be taken into account.
In this paper, Gordon effective metric is introduced to deal with the interaction between
the emission and the gravitation, as well as the electromagnetic interaction between the emission
and the medium. Namely, the gravitational and the electromagnetic effects are together geometrized
in a unified effective metric, and the standard geometric procedure used in general relativity
to describe the photons can now be used upon replacing the gravitational metric by the effective metric.

Gordon metric is deduced from Maxwell's equations in a curved space-time, which is defined as \citep{GO23,EH67}
\begin{equation}
G_{\mu\nu}=g_{\mu\nu}+\left(\frac{1}{n^2}-1\right)u_\mu u_\nu,
\label{4}
\end{equation}
where $g_{\mu\nu}$ is the gravitational metric which can be reduced to Minkowski metric
$\eta_{\mu\nu}=\textmd{diag}(1,-1,-1,-1)$ when no gravity, $n$ is the refractive index of the medium, and
$u_\mu$ is the four-vector velocity of the medium denoted as $u_\mu=\gamma(1,\textbf{\emph{V}}/c)$
($\textbf{\emph{V}}$ is the three-dimensional velocity of medium seen by a distant static observer,
with the Lorentz factor $\gamma=(1-\beta^2)^{-1/2}$, where $\beta=V/c$).
It should be pointed out that Gordon metric was presented originally for the light propagating in a
moving non-dispersive medium. However, in the case of the susceptibility of the medium $|\chi|\ll1$,
the Gordon metric can be generalized to a weak-dispersive medium (see Appendix in detail).
Light traveling in a dispersive medium can be described equivalently as light propagating
in a curved space-time characterized by an effective metric under an external field specified by an
effective potential $A^\mu$. In such case, the kinetic momentum of a photon, $dx^{\mu}/d\lambda=(\omega/c,\textbf{\emph{k}})$,
is no longer following null geodesics due to the dispersive effect of the medium,
while the canonical momentum, $K^{\mu}=dx^{\mu}/d\lambda-A^{\mu}=(\omega/c-A^0,\textbf{\emph{k}}-\textbf{\emph{A}}$),
is governed by the geodesics equation,
where $\lambda$ is the affine parameter along the ray, $\textbf{\emph{k}}$ and $\textbf{\emph{A}}$ are
three-dimensional wave vector and effective potential respectively. Thus the Gordon metric is still valid
as an effective metric in the cases of light propagating in a weak-dispersive medium, but it is required
that an effective potential is added because of the dispersive effect.

The redshift of the spectrum contributed by the effects of the gravitational field and the medium
around pulsars is now corrected as
\begin{equation}
1+\tilde{z}=\frac{\Omega_e}{\Omega_o}=\sqrt{\frac{G_{00}(o)}{G_{00}(e)}},
\label{5}
\end{equation}
where $G_{00}(e)$ and $G_{00}(o)$ denote the time-time components of Gordon metrics at the emitter
and the observer respectively, $\Omega_e$ and $\Omega_o$ are the time-time components of canonical
momenta of photons at the emitter and the observer respectively, called the canonical frequencies,
and $\tilde{z}$ is the redshift for the canonical frequency.
The above formula (\ref{5}) can be further written as
\begin{equation}
\frac{\nu_e-\frac{cA^0}{2\pi}}{\nu_o}=1+z-\frac{cA^0}{2\pi\nu_0}=\sqrt{\frac{G_{00}(o)}{G_{00}(e)}}.
\label{6}
\end{equation}
Because of the effective potential $A^0=\chi'(u^{\mu}k_{\mu})^2u^0/2\sim|\chi|/\lambda_\rmn{w}$
($\lambda_\rmn{w}$ is the wavelength of the photon), we can obtain
\begin{equation}
1+z\simeq\left[g_{00}(e)+\left(\frac{1}{n^2}-1\right)u_0u_0\right]^{-1/2}+\frac{|\chi|}{2\pi}.
\label{7}
\end{equation}
Considered that $n^2=1+\chi$ ($|\chi|\ll1$) and $u_0=\gamma$, Eq. (\ref{7}) can be reduced to
\begin{equation}
1+z\simeq\left[1-\frac{2GM}{c^2R}-\chi \gamma^2\right]^{-1/2}+\frac{|\chi|}{2\pi}.
\label{8}
\end{equation}
Expanding the first term in the right side of Eq. (\ref{8}), we can obtain
$z\sim GM/(c^2R)+\chi\gamma^2/2+|\chi|/(2\pi)$. The third term $|\chi|/(2\pi)$
is much smaller than the first two terms, so Eq. (\ref{8}) can be approximated as
\begin{equation}
1+z\simeq\left[1-\frac{2GM}{c^2R}-\chi \gamma^2\right]^{-1/2}.
\label{9}
\end{equation}
In this case, the modified redshift is now proven to have a couple of components:
one coming from the gravitational effect, and another from the electromagnetic and dynamic effects of the medium which
is manifested by the dispersion relation and the four-vector velocity $u_\mu=\gamma(1,\textbf{\emph{V}}/c)$ of
the streaming medium. Here $\textbf{\emph{V}}$ is dependent on the motion and rotation of pulsars and
the streaming of the plasma. Observations show that mean space velocity of pulsars is
$v_\rmn{space}=450\pm50$ $\textmd{km}$ $\textmd{s}^{-1}$ \citep{LY94}. In a corotation medium frame,
the rotation velocity of plasma is estimated as $v_\rmn{rotaion}=2\pi R/P\approx62.8$ $\textmd{km}$ $\textmd{s}^{-1}$
if $R\sim10$ km and $P=1$ s. The resultant velocity of pulsars is about $v_\rmn{resultant}\sim10^3$ $\textmd{km}$ $\textmd{s}^{-1}$
which is very small compared to the streaming velocity of pair plasma, and can be ignored.
Finally, the redshift of the spectrum caused by the magnetized plasma and the vacuum polarization is
\begin{equation}
\delta\approx\left[1-\frac{2GM}{c^2R}-\chi \gamma^2\right]^{-1/2}-\left[1-\frac{2GM}{c^2R}\right]^{-1/2}.
\label{10}
\end{equation}
It can be seen from Eq. (\ref{10}) that the magnitude of the modification of redshift is approximated to
$\sim\chi\gamma^2/2$. If $\chi>0$ and $\delta>0$, the gravitational redshift can be strengthened by
the effect of the medium. Conversely, if $\chi<0$ and $\delta<0$, the gravitational redshift is weakened.

The weak-dispersive condition is generally satisfied in the regime of magnetized pair plasmas of pulsars
and magnetars, except for a singular point in the dispersion relation of electromagnetic emissions.
Thus, in this paper, we will focus on the emissions with frequencies being in the range of
$\omega_{p}/\sqrt{\gamma}\ll\omega\ll\omega_{c}/\gamma$ and $\omega\gg\omega_{c}/\gamma$,
in which the weak-dispersive condition is satisfied and the Gordon metric can be used to describe the
propagation of emissions.
Here $\omega_{p}=(4\pi Ne^2/m_{e})^{1/2}\simeq1.49\times10^{10}\eta^{1/2} B_{12}^{1/2}P_1^{-1/2}$ $\textmd{s}^{-1}$
and $\omega_{c}=eB/m_{e}c\simeq1.76\times10^{19}B_{12}$ $\textmd{s}^{-1}$ are non-relativistic
plasma frequency and cyclotron frequency of electron and positron respectively.
Pair cascade simulation generally give the number density of the pair plasma $N=\eta N_\rmn{GJ}$,
where $\eta\sim10^2-10^5$ is multiplicity factor for the pulsar \citep{DA82,HI01,ME10,HA11},
and $N_\rmn{GJ}=B/Pec\simeq7.0\times10^{10}B_{12}P_1^{-1}$ $\textmd{cm}^3$ is the Goldreich-Julian
number density ($B_{12}=B/(10^{12}\textmd{G})$, $P_1=P/(1\textmd{s})$,
$B$ and $P$ is the local magnetic field and the period of pulsars, respectively).

\section{Calculation of the modification of redshift}
When electromagnetic radiations pass though the magnetized pair plasma of pulsars, the redshifts
of radiation spectra will be changed. Furthermore, it has long been predicted from the quantum
electrodynamics (QED) that the vacuum becomes birefringent in a strong magnetic field. When combined
with the birefringence due to the magnetized plasma, vacuum polarization can greatly affect the
propagation of radiations. In this section, we will calculate the modification of redshifts of spectral
lines attributed to the plasmas around pulsars and magnetars and the QED vacuum polarization effect
caused by the super-strong magnetic field of stars.

\subsection{Redshift modified by magnetized plasmas}
The magnetospheres of pulsars and magnetars consist of relativistic electron-positron pairs streaming along
magnetic field lines \citep{ST71,CH86,RO96,ZH00,HI01,TH02,BE07}.
Wave modes in pulsar magnetospheres have been studied in a number of papers under
different assumptions about the plasma composition and the velocity distribution of electron-positron pairs \citep{ME77,AR86,LY98,ME99,AS00}.
When the angle $\theta$ between the wave vector $\textbf{\emph{k}}$ and the magnetic field
$\textbf{\emph{B}}$ is not close to $0^\circ$ or $180^\circ$,
electromagnetic waves propagate in two normal modes in the magnetized plasma \citep{ME77}.
The ordinary mode (O-mode) is mostly polarized parallel to the $\textbf{\emph{k}}-\textbf{\emph{B}}$ plane,
and the extraordinary mode (X-mode) is mostly polarized perpendicular to the $\textbf{\emph{k}}-\textbf{\emph{B}}$ plane.

We consider a cold electron-positron plasma with all the electrons and positrons having the same velocity
($\textbf{\emph{V}}_-=\textbf{\emph{V}}_+, \beta_-=\beta_+=\beta, \gamma_-=\gamma_+=\gamma$).
For the radiations with frequencies being in the range of $\omega_{p}/\sqrt{\gamma}\ll\omega\ll\omega_{c}/\gamma$,
the index of refraction for the O-mode is \citep{WA07}
\begin{equation}
n^2=1+\chi\simeq1+f_{\eta}\sin^2\theta+f_{11},
\label{11}
\end{equation}
and for the X-mode is
\begin{equation}
n^2=1+\chi\simeq1+f_{11},
\label{12}
\end{equation}
where
\begin{equation}
f_{\eta}\simeq-v\gamma^{-3}(1-n\beta\cos\theta)^{-2},
\label{13}
\end{equation}
\begin{equation}
f_{11}\simeq vu^{-1}\gamma(1-n\beta\cos\theta)^{2}
\label{14}
\end{equation}
with $v=\omega_{p}^2/\omega^2$ and $u=\omega_{c}^2/\omega^2$.
In such frequency range, $|f_{\eta}|\ll1$ and $|f_{11}|\ll1$,
so the weak-dispersive condition is satisfied.

The modification of redshifts by magnetized pair plasmas is originated from two factors: the electromagnetic effect of
the plasma manifested by the refractive index, and the kinetic effect of the plasma manifested by the four-vector velocity
$u_\mu=\gamma(1,\textbf{\emph{V}}/c)$.
From Eqs. (\ref{11}) and (\ref{12}), we can obtain the susceptibility $\chi$ of the plasma medium.
And then inserting $\chi$ into Eq. (\ref{10}), we can get the modification of redshifts for the two modes
due to the magnetized plasma.
Because of $|\chi|\ll1$, we can simply set $n\approx1$ in Eqs. (\ref{13}) and (\ref{14}) for estimating
the order of magnitude of the redshift, which does not affect the calculation.
The susceptibilities of magnetized plasmas for the two modes can be approximated to
\begin{equation}
\chi_{p-O}\approx-v\gamma^{-3}(1-\beta\cos\theta)^{-2}\sin\theta^2+vu^{-1}\gamma(1-\beta\cos\theta)^{2},
\label{15}
\end{equation}
\begin{equation}
\chi_{p-X}\approx vu^{-1}\gamma(1-\beta\cos\theta)^{2},
\label{16}
\end{equation}
and the order of the modification is $\delta_p\sim\chi_{p}\gamma^2/2$.

In pulsars, the Lorentz factor of bulk movement of the plasma flowing outward along magnetic field lines is
$\gamma\sim10^2-10^4$ \citep{DA82,HI01,ME10,HA11}.  From Eqs. (\ref{10}) - (\ref{12}), we can depict
the dependences of the susceptibility $\chi_p$ and the modification of redshift $\delta_p$ on the
wave frequency of pulsars in the range of $\omega_p/\sqrt{\gamma}\ll\omega\ll\omega_c/\gamma$.
Fig. 1 is the case of radio radiation of the pulsar. Radio radiation is thought to originate in the
pair plasma at a distant of $r\sim100$ $\textmd{km}\approx10R$ from the center of stars \citep{CO78}.
If the surface magnetic field of a pulsar is $B_\ast=10^{12}$ $\textmd{G}$, the local magnetic field
in this region is $B\simeq B_\ast(R/r)^3\sim10^9\textmd{ G}$ using the dipole magnetic field approximation.
Fig. 2 is the case of typical local magnetic field of the pulsar, if emissions happen near the surface of
the pulsar. For the O-mode, the effect of magnetized plasmas is to weaken the redshift of the spectrum
($\delta_p<0$). The modification of the redshift depends on the wave frequency. For some lower frequencies,
the magnitude $|\delta_p|\sim10^{-1}$ can be the same scale as the gravitational redshift. For the X-mode,
the redshift of the spectrum is strengthened by the effect of magnetized plasma ($\delta_p>0$).
The modification is irrelevant to the wave frequency and much smaller than the case of O-mode for the same parameters.

For magnetars, theoretical works suggest that a corona consists mainly of relativistic pairs with $\gamma\sim10^3$
(and a wide spread in $\gamma$) \citep{TH02,BE07}. The multiplicity factor of the pair plasma is of the order of
$\eta\sim2\times10^3(R/r)$. Moreover, observed radiations are usually in the X-ray band. Assuming the magnetars
with parameters $B\simeq10^{14}-10^{15}$ G, $P\sim10$ s, $\omega\sim10^{18} \textmd{s}^{-1}$, $\gamma\sim10-10^3$ and
$\eta\sim10^3$, we have $|\chi_{p-O}|\sim|\chi_{p-X}|<10^{-16}$ and $|\delta_{p}|<10^{-10}$.
Obviously, the modification of redshifts by plasmas can be neglected for the X-rays of magnetars.

\begin{figure}
\centering
\begin{minipage}[c]{0.5\textwidth}
\includegraphics[width=9.5cm]{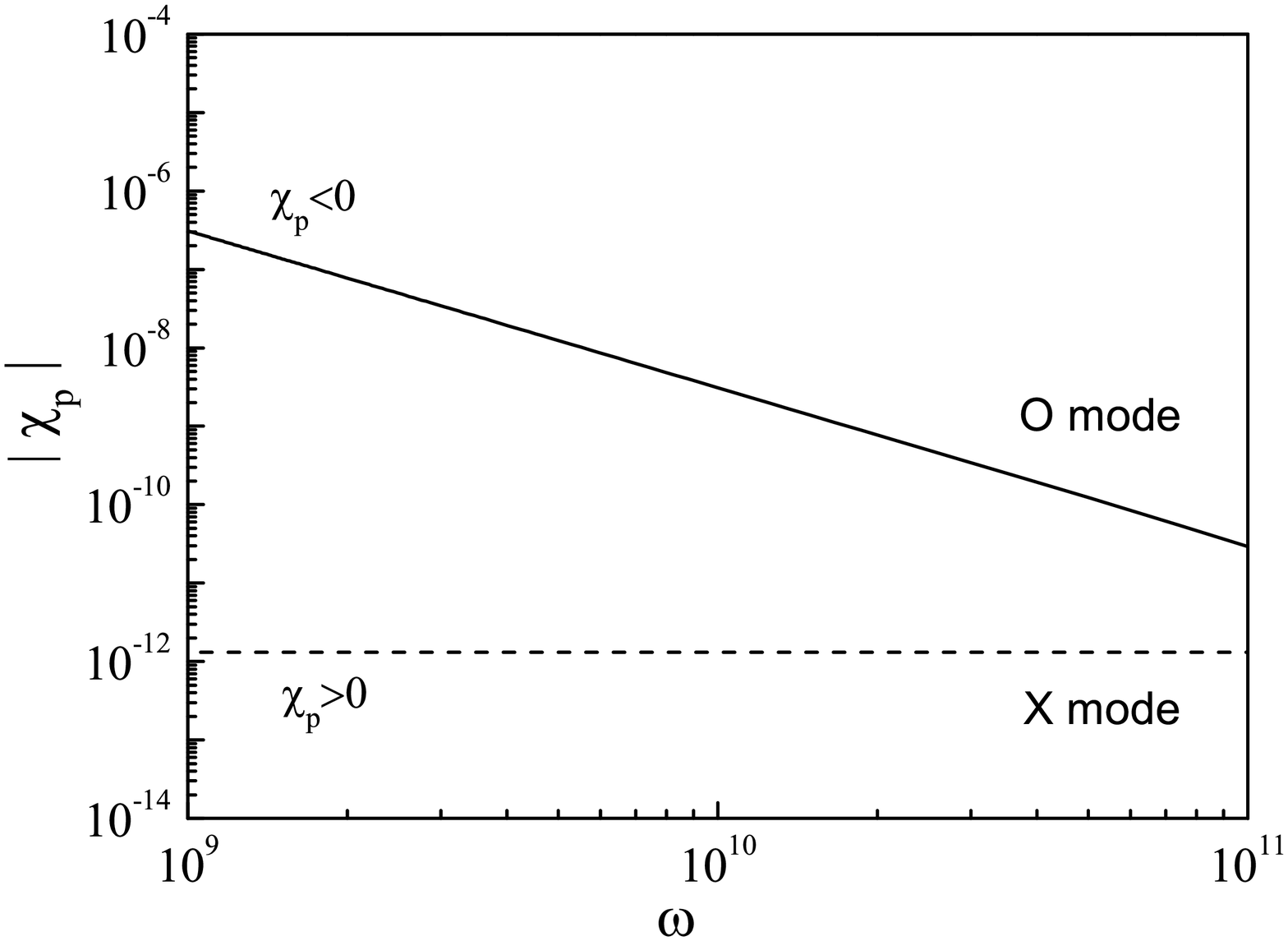}
\end{minipage}%
\begin{minipage}[c]{0.5\textwidth}
\includegraphics[width=9.5cm]{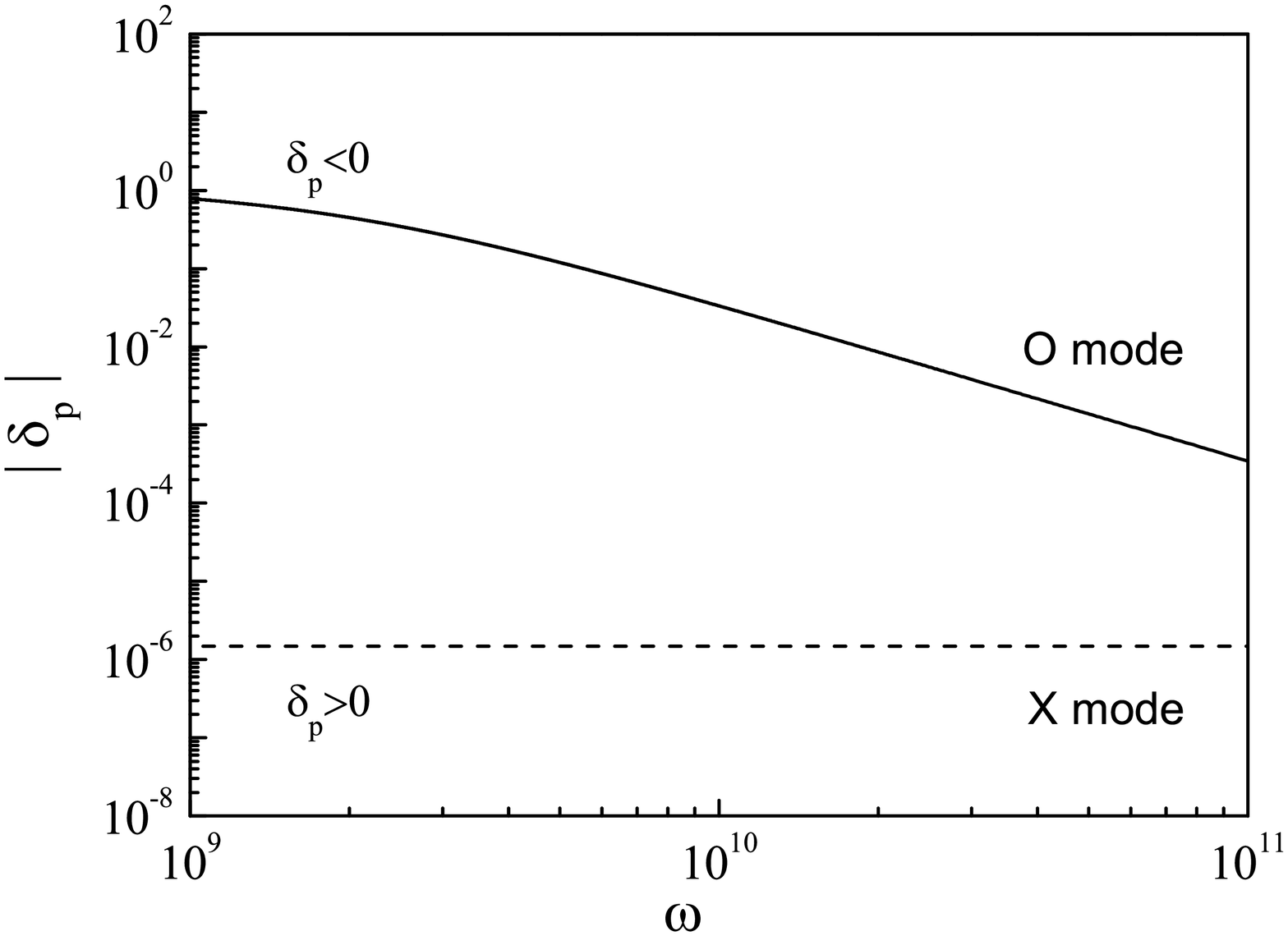}
\end{minipage}
\caption{The dependences of the susceptibility and the modification of redshift, caused by magnetized pair plasmas, on the frequency.
The case of the O-mode is represented by the solid line, and the X-mode is represented by the dashed line.
The parameters are $B=10^{9}\textmd{G}$, $P=1\textmd{s}$, $\eta=10^2$, $\gamma=10^3$ and $\theta=30^\circ$. }
\end{figure}

\begin{figure}
\centering
\begin{minipage}[c]{0.5\textwidth}
\includegraphics[width=9.5cm]{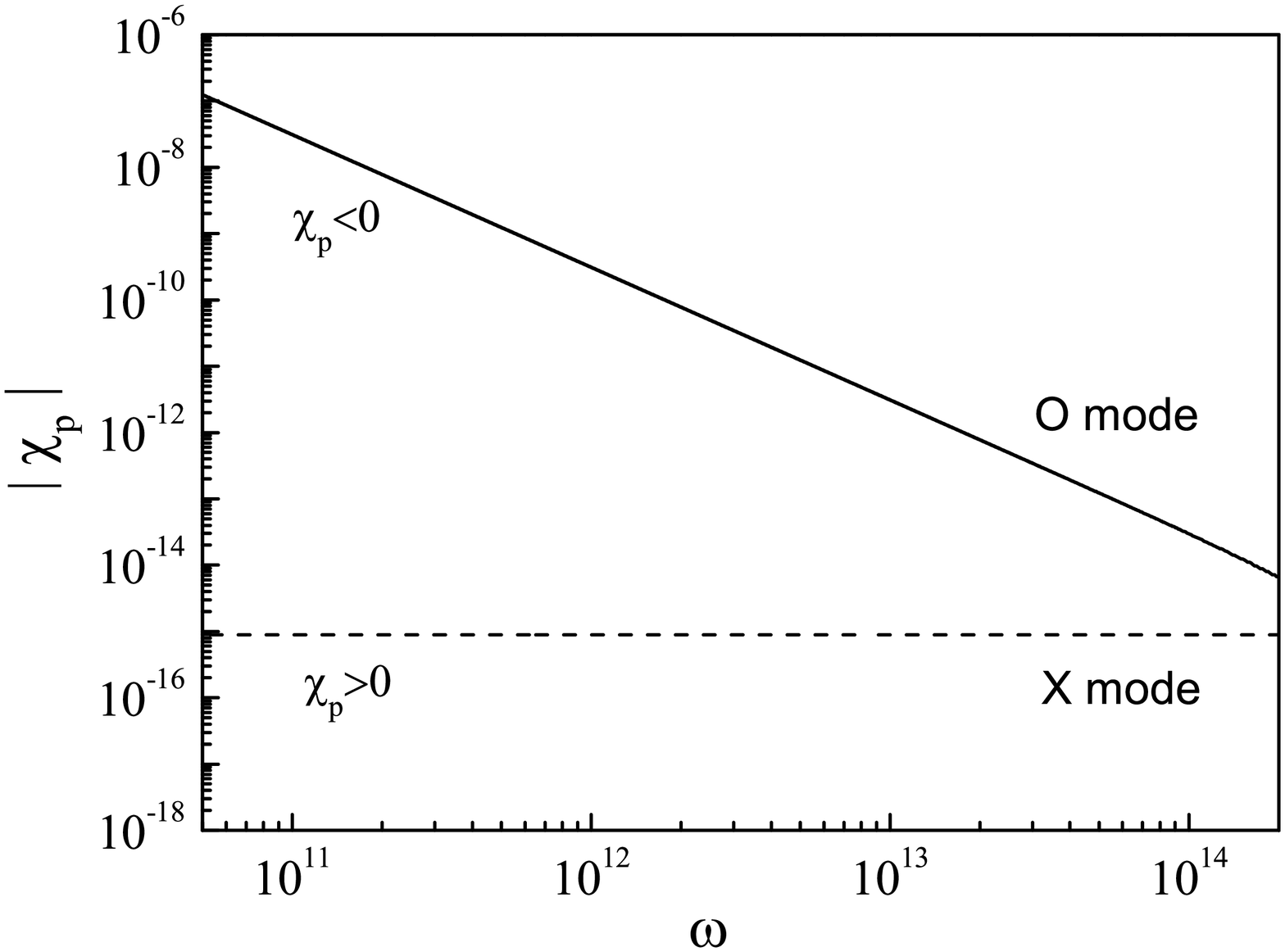}
\end{minipage}%
\begin{minipage}[c]{0.5\textwidth}
\includegraphics[width=9.5cm]{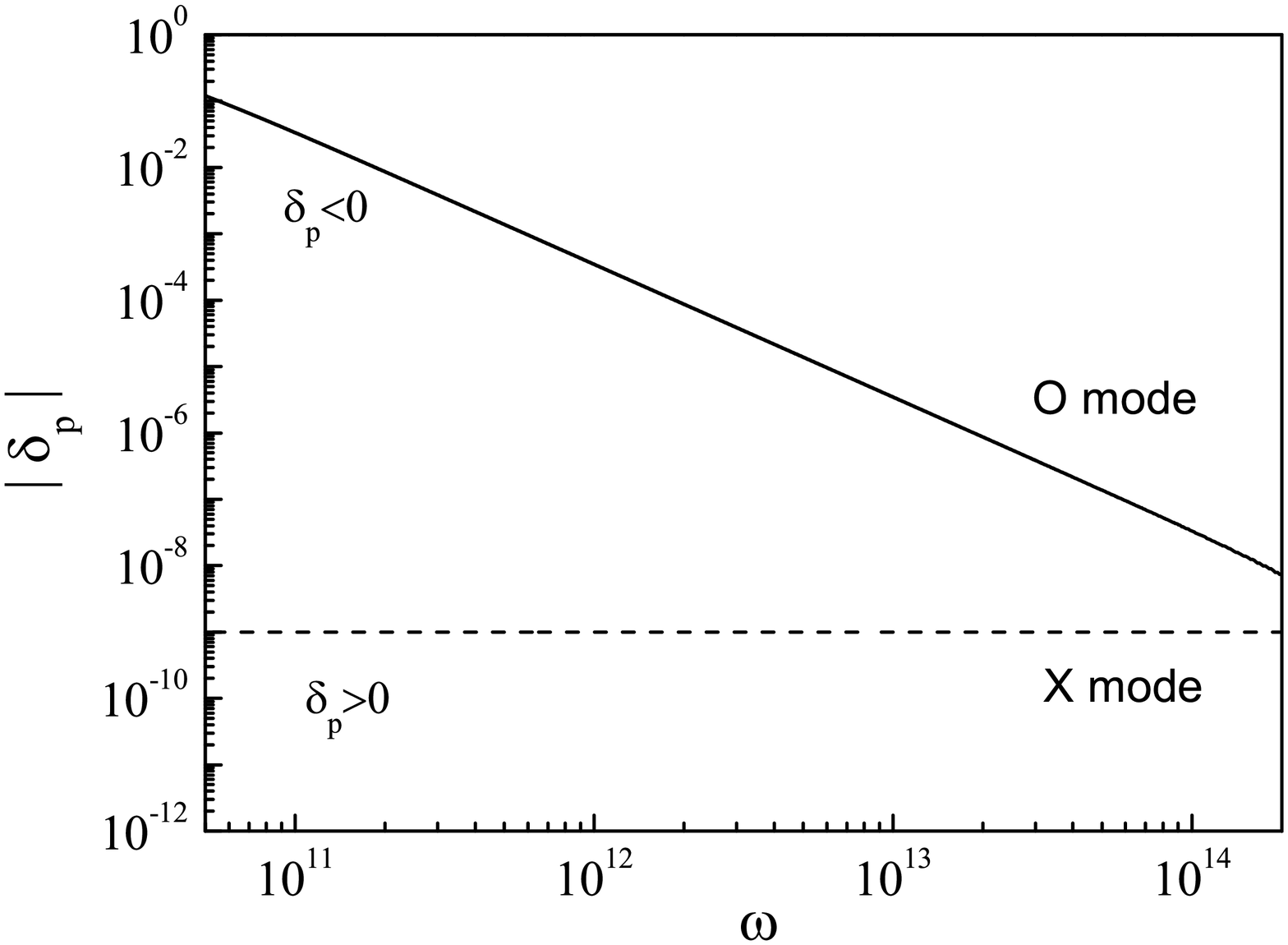}
\end{minipage}
\caption{The dependences of the susceptibility and the modification of redshift, caused by magnetized pair plasmas, on the frequency.
The parameters are $B=10^{12}\textmd{G}$, $P=1\textmd{s}$, $\eta=10^2$, $\gamma=10^3$ and $\theta=30^\circ$.}
\end{figure}

For the radiations with higher frequencies $\omega\gg\omega_c/\gamma$,
the functions $f_{11}\simeq-v\gamma^{-1}$ and $|f_\eta|\approx v\gamma^{-3}\ll|f_{11}|\ll1$ \citep{WA07}.
The refractive indices for the O-mode and X-mode are expressed as
\begin{equation}
n_O^2\approx n_X^2\approx1-v\gamma^{-1}=1-\omega_p^2/\gamma\omega^2.
\label{17}
\end{equation}
In such case, the magnitude of the redshift modification is $|\delta_p|\ll\gamma^3\omega_p^2/2\omega_c^2$.
For typical parameters of pulsars, $B=10^{12}$ $\textmd{G}$, $P=1$ $\textmd{s}$, $\eta\simeq10^2-10^5$,
$\gamma\sim10^2-10^4$ and $\theta=30^\circ$, we can obtain $\chi_{p-O}\simeq\chi_{p-X}\ll10^{-10}$
and $|\delta_p|\ll10^{-2}$.
Thus the modification of redshift for the higher-frequency radiation can be neglected in general case.

\subsection{Redshift modified by magnetized plasmas and vacuum polarization}
In the regime of pulsars and magnetars, the vacuum becomes birefringent in super-strong magnetic fields, and
can greatly affect the propagation of radiations.
Therefore, the non-gravitational redshifts caused by the QED vacuum polarization effect
should also be taken into consideration

We now consider how vacuum polarization affects the redshifts of spectral lines.
The wave modes propagating in a combined `plasma+vacuum' medium
have been studied \citep{KI80,WA07}.
In a homogeneous pair plasma, there is a special wave frequency, called ``vacuum resonance frequency'',
due to the combined effects of vacuum polarization and magnetized plasmas \citep{WA07}.
The frequency is expressed as, at given $B$, $\gamma$ and density $N$,
\begin{equation}
\omega_v=1.67\times10^9[\eta B_{12}^{-1}P_1^{-1}\gamma_3^{-3}(1-n\beta\cos\theta)^{-2}F^{-1}]^{1/2} \textmd{s}^{-1}
\label{18}
\end{equation}
where $\gamma_3=\gamma/10^3$, $F=(q+m)/(3\zeta_v)$, $\zeta_v=\alpha/(45\pi)(B/B_Q)^2=2.650\times10^{-8}B_{12}^2$,
$q$ and $m$ are functions of $B$ \citep{AL71,HE97},
and $\alpha=e^2/\hbar c=1/137$ is the fine structure constant.
Theoretical studies showed that wave modes with $\omega\ll\omega_v$ are determined by the plasma effect, and those with $\omega\gg\omega_v$
are determined by the vacuum polarization effect.
When the wave frequency is not very close to the ``vacuum resonance frequency'', the wave can still be described by normal modes.
The refractive index for the O-mode is
\begin{equation}
n^2=1+\chi\simeq1+q\sin^2\theta+f_\eta\sin^2\theta+f_{11},
\label{19}
\end{equation}
and for the X-mode is
\begin{equation}
n^2=1+\chi\simeq1+m\sin^2\theta+f_{11},
\label{20}
\end{equation}
where the terms $q\sin^2\theta$ and $m\sin^2\theta$ are originated from the vacuum polarization effect.
We can separate the susceptibility into two parts, i.e. $\chi=\chi_p+\chi_v$, according to the contributions of different effects.
Here, the susceptibilities $\chi_p$ contributed by magnetized plasmas for the two modes can be expressed by Eqs. (\ref{15}) and (\ref{16}),
and the susceptibilities $\chi_v$ contributed by the vacuum polarization effect can be written as
\begin{equation}
\chi_{v-O}\simeq q\sin^2\theta,
\label{21}
\end{equation}
\begin{equation}
\chi_{v-X}\simeq m\sin^2\theta.
\label{22}
\end{equation}

First, we still consider the radiations with frequencies $\omega_p/\sqrt{\gamma}\ll\omega\ll\omega_c/\gamma$.
For typical pulsars, $B\ll B_Q=m_e^2c^3/e\hbar=4.414\times10^{13}$ $\textmd{G}$,
the parameters $q=7\zeta_v$ and $m=4\zeta_v$ \citep{AL71}.
From Eqs. (\ref{19}) and (\ref{20}), the dependences of the susceptibility and the modification of redshifts on
the frequency are plotted in Fig. 3 and Fig. 4. The case of radio radiation of the pulsar is depicted in Fig. 3.
It can be seen that the modification of redshifts by magnetized plasmas and vacuum polarization is entirely the
same as the case of only considering the plasma effect (see Fig. 1). The reason is that the magnetic field in the
region of radio radiation is weaker, so the vacuum polarization effect can be ignored and the wave property is
dominated by the plasma effect.

An interesting phenomenon exists when the ``vacuum resonance frequency'' is in the range of
$\omega_p/\sqrt{\gamma}\ll\omega\ll\omega_c/\gamma$ (see Fig. 4). The circle in Fig. 4 denotes the region
of the ``vacuum resonance frequency''.
It can be seen that the effects of the medium on the radiation are completely different for the two cases:
$\omega<\omega_v$ and $\omega>\omega_v$.
The magnitude of the modification for the X-mode can reach $\delta\approx10^{-2}$ for the typical parameters of pulsars.
The modification becomes much larger than the case of only considering the plasma effect, and
is mainly attributed to the vacuum polarization effect.
Moreover, the redshift for the X-mode is strengthened by the vacuum polarization effect ($\delta>0$).
For the O-mode,  the situation is somewhat complicated.
When $\omega>\omega_v$, the redshift of radiations is strengthened and dominated by the vacuum polarization effect.
When $\omega<\omega_v$, there is a special frequency $\omega_0$ satisfying the equation $q\sin^2\theta+f_\eta\sin^2\theta+f_{11}=0$,
at which $\chi=0$ and thus there is no modification to the redshift. This is the result of the combined effects of plasma
and vacuum polarization. Therefore, for the radiation with this frequency $\omega_0$, the redshift is a pure gravitational effect.
In the case of $\omega<\omega_0$, the modification of the redshift for the O-mode is weakened ($\delta<0$) and attributed to the magnetized plasma.
While in the case of $\omega>\omega_0$, the modification is strengthened ($\delta>0$) by the vacuum polarization effect.
Generally speaking, the  modification of redshift for the O-mode should be taken into account, except for a very small range of frequencies near $\omega_0$.

\begin{figure}
\centering
\begin{minipage}[c]{0.5\textwidth}
\includegraphics[width=9.5cm]{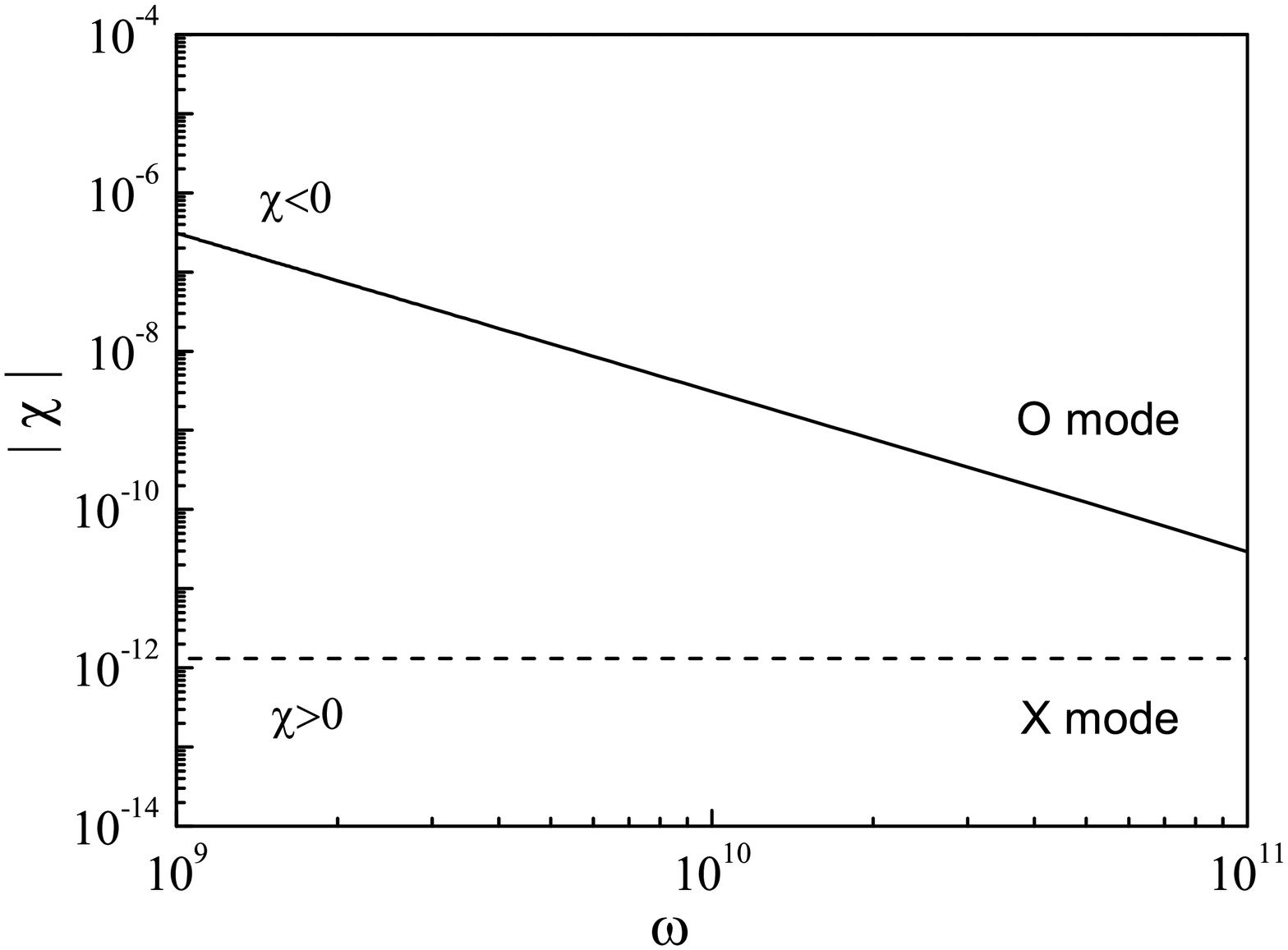}
\end{minipage}%
\begin{minipage}[c]{0.5\textwidth}
\includegraphics[width=9.5cm]{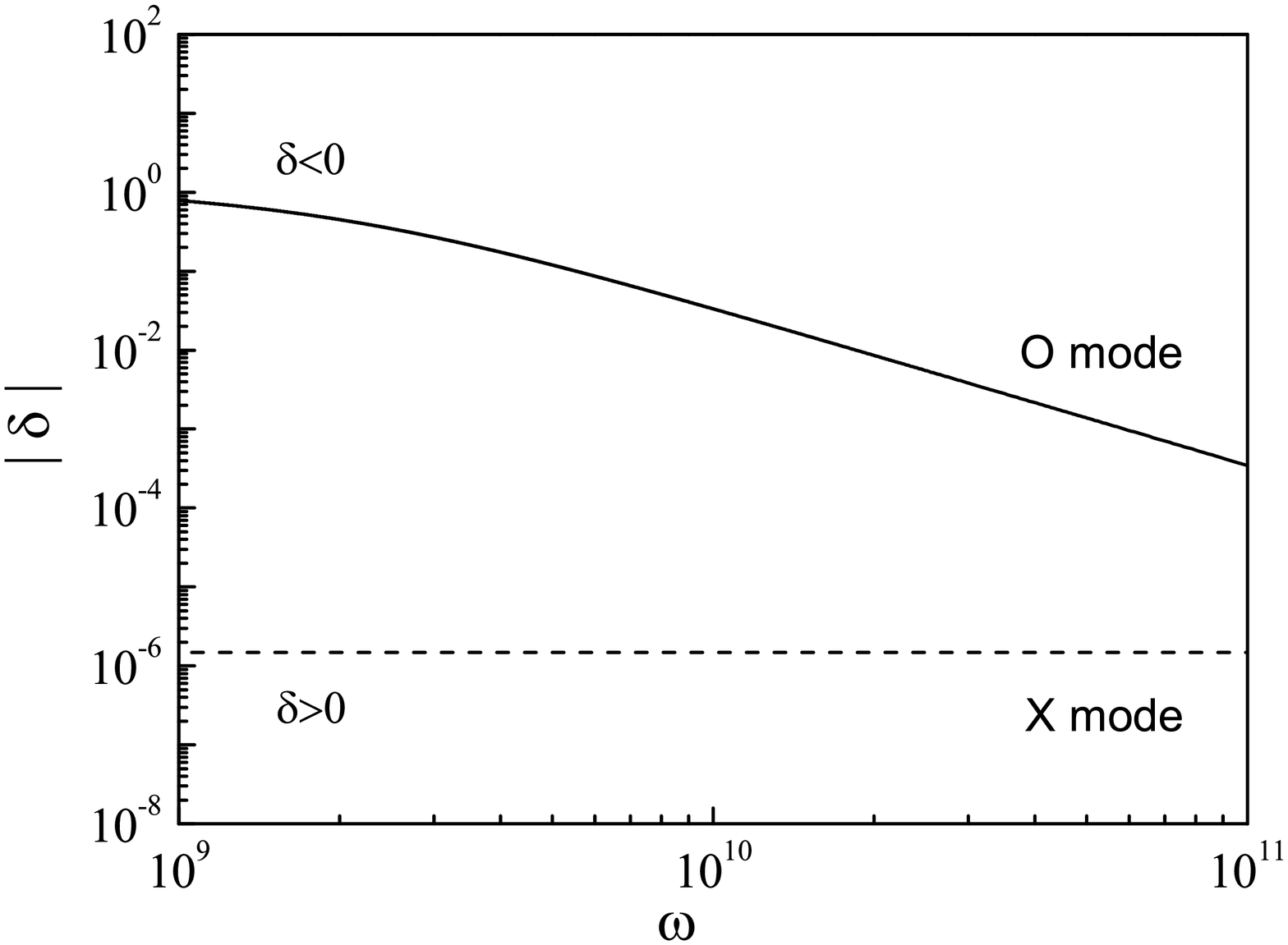}
\end{minipage}
\caption{The dependences of the susceptibility and the modification of redshift, caused by magnetized pair plasmas, on the frequency.
The parameters are $B=10^{9}\textmd{G}$, $P=1\textmd{s}$, $\eta=10^2$, $\gamma=10^3$ and $\theta=30^\circ$.}
\end{figure}

\begin{figure}
\centering
\includegraphics[width=12cm]{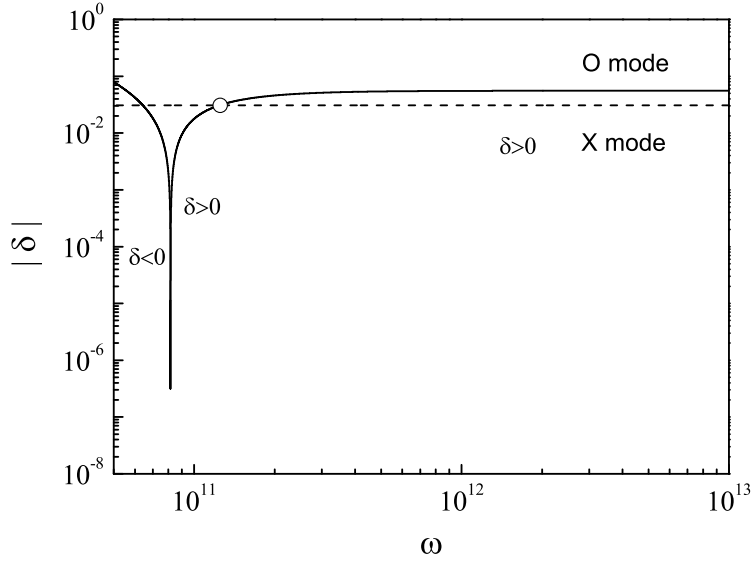}
\caption{The dependences of the modification of redshift, caused by magnetized pair plasmas and vacuum polarization, on the frequency.
The circle denotes the region of the ``vacuum resonance frequency''. The parameters are $B=10^{12}\textmd{G}$,
$P=1\textmd{s}$, $\eta=10^2$, $\gamma=10^3$ and $\theta=30^\circ$.}
\end{figure}

For magnetars, the magnetic fields $B>B_Q$, so
the parameters $q$ and $m$ are, using the expansion by \citet{HE97},
\begin{equation}
q\simeq-\frac{\alpha}{2\pi}\left[-\frac{2}{3}b+1.272-\frac{1}{b}(0.307+\ln b)-\frac{0.7}{b^2}\right],
\label{29}
\end{equation}
\begin{equation}
m\simeq\frac{\alpha}{2\pi}\left[\frac{2}{3}+\frac{1}{b}(0.145-\ln b)-\frac{1}{b^2}\right],
\label{30}
\end{equation}
where $b=B/B_Q$.
Assuming the magnetic fields of magnetars $B\simeq10^{14}-10^{15}$ G, we can obtain the values
$q\sim8\times10^{-4}-1.6\times10^{-2}$ and $m\sim2\times10^{-4}-6\times10^{-4}$,
as well as $\chi_{v-O}\sim6\times10^{-4}-1.2\times10^{-2}$ and $\chi_{v-X}\sim1.5\times10^{-4}-4.5\times10^{-4}$
($\theta=30^\circ$).
Obviously, $\chi_{v}\gg\chi_{p}$ for the same parameters in Section 3.1.
In this case, the refractive indices for the two modes can be approximated to
\begin{equation}
n_O^2\simeq1+q\sin^2\theta,
\label{35}
\end{equation}
\begin{equation}
n_X^2\simeq1+m\sin^2\theta,
\label{36}
\end{equation}
and the modification of the redshift is $\delta\sim\chi_v\gamma^2/2$.
We can see that the vacuum polarization effect dominates the propagation property of X-ray radiations of magnetars,
and the modification of the redshift should be taking into consideration as long as the Lorentz factor of streaming plasma
$\gamma>10$.

Now, we discuss the case of frequencies being in the range of $\omega\gg\omega_c/\gamma$.
Using the same pulsar parameters as in Section 3.1, we obtain $\chi_{v-O}\sim1.4\times10^{-7}$ and $\chi_{v-X}\sim8\times10^{-8}$.
Because of $\chi_{v}\gg\chi_{p}$, the refractive indices can also be expressed as Eqs. (\ref{35}) and (\ref{36}).
In such case, the modification of the redshift is independent of the wave frequency, and the magnitudes are
$\delta_O\sim7\times10^{-2}$ and $\delta_X\sim4\times10^{-2}$ assuming $\gamma\sim10^3$.
Compared to the case of only considering the plasma effect, the modification increases several orders of
magnitude, and is attributed to the vacuum polarization effect.

\section{Discussion and conclusion}
In astrophysics, the gravitational redshift is used to infer the mass-radius relation, and thus
the equation of state of a compact star. In this paper, the modification of the gravitational
redshift caused by magnetized pair plasmas and vacuum polarization
in pulsars and magnetars are investigated within the framework of Gordon effective metric.
The Gordon's effective metric theory is generalized to the weak-dispersive media.
The propagation of photons in magnetized pair plasmas and vacuum can be described as
if propagating in a curved space-time characterized by an effective metric under an external field specified
by an effective potential. The gravitational metric $g_{\mu\nu}$ is replaced by the effective metric $G_{\mu\nu}$.
In this case, the redshift has two contributions: one coming from gravitational effects and another
from the effects of the medium.

The modification of the gravitational redshift due to the magnetized plasma and vacuum polarization is discussed in detail.
We find that the magnetized plasmas and vacuum polarization play an important role in the propagation of radiations.
The modification of redshift can be the same magnitude as the gravitational redshift under certain conditions.
In this case, we should be careful when using the gravitational redshift to infer the mass-radius ratio of the observed pulsar.
Unlike the gravitational redshift being isotropic, the redshift caused by the plasmas and vacuum polarization is related to the wave modes.
The redshift of O-mode heavily depends on the frequencies of waves in the range of $\omega_p/\sqrt{\gamma}\ll\omega\ll\omega_c/\gamma$.
In a combined `plasma+vacuum' medium, there is a special frequency $\omega_v$ due to the effects of magnetized plasmas
and vacuum polarization. For $\omega\ll\omega_v$, the modification of redshifts is determined by the plasma effect,
while for $\omega\gg\omega_v$, it is determined by the vacuum polarization effect.
There is also a special frequency $\omega_0$, at which the plasma effect and vacuum polarization
effect cancel out each other, so the susceptibility $\chi=0$. In this case, the redshift of the O-mode with this
frequency is a pure gravitational effect.
The redshift of the X-mode is non-dispersive, and its modification of redshift is mainly attributed to the vacuum
polarization effect.
For the radiations with higher frequencies, i.e. $\omega\gg\omega_c/\gamma$, the modification for the two modes
are all approximately non-dispersive, and attributed to the vacuum polarization effect.
For magnetars, the modification of the redshift of X-rays is also dominated by the vacuum polarization effect,
and may overtake the gravitational redshift.

At present, the observed absorption lines are usually in the X-ray band. According to the discussion in this paper,
the modification of the gravitational redshift is attributed to the vacuum polarization effect.
Surface emission from pulsars and magnetars provides a useful probe for interior physics, surface magnetic
fields, and composition of stars. The frequency of thermal radiation is in the soft X-ray band,
so the modification is also due to the vacuum polarization effect.
The magnitude of the modification depends on characters of plasmas in the magnetospheres of pulsars and magnetars,
such as the streaming velocity and the distribution of particles.
So we hope that exact numerical results of redshifts caused by the media can be obtained when
the plasma parameters are confirmed further.

\section*{Acknowledgments}
This work was partly supported by the Shanghai Leading
Academic Discipline Project (Project No. S30105) and the Shanghai
Research Foundation (Grant No. 07dz22020).

\appendix
\section{The generalization of Gordon metric}

It is well known that the effective metric $G_{\mu\nu}=g_{\mu\nu}+(1/n^{2}-1)u_{\mu}u_{\nu}$
was presented originally by Gordon for the light propagating in a moving non-dispersive medium.
From the ray equations
\begin{equation}
\frac{dx^{\mu}}{d\lambda}=\frac{\partial\mathcal{H}}{\partial k_{\mu}},
\end{equation}
\begin{equation}
\frac{dk_{\mu}}{d\lambda}=-\frac{\partial\mathcal{H}}{\partial x^{\mu}},
\end{equation}
and the dispersion relation for waves $\mathcal{H}(x^{\mu},k_{\nu})=0$,
one can obtain the null geodesic equation for the propagation of light in a moving non-dispersive medium.
Here $k_{\mu}$ is the four-dimensional wave vector, $x^{\mu}$ is the canonical variable,
$\lambda$ is the affine parameter along the ray, $\mathcal{H}=\frac{1}{2}G^{\mu\nu}k_{\mu}k_{\nu}$
is interpreted as the Hamiltonian for the photon and $G^{\mu\nu}=G^{\mu\nu}(x)=g^{\mu\nu}+(n^2-1)u^{\mu}u^{\nu}$
is an effective metric for the non-dispersive medium.
Inserting the Hamiltonian $\mathcal{H}$ to the ray equation (A1), one can obtain
\begin{equation}
\frac{dx^{\mu}}{d\lambda}=G^{\mu\nu}k_{\nu}=K^{\mu}.
\end{equation}
Using the effective metric $G^{\mu\nu}$, the four-vector $k_{\mu}$ can  be expressed as $k_{\mu}=G_{\mu\nu}K^{\nu}$.
Then from Eqs. (A2) and (A3), we can
obtain the null geodesic equation
\begin{equation}
\frac{d^{2}x^{\mu}}{d\lambda^{2}}+\Gamma^{\mu}_{\nu\gamma}\frac{dx^{\nu}}{d\lambda}\frac{dx^{\gamma}}{d\lambda}=0,
\end{equation}
where
\begin{equation}
\Gamma^{\mu}_{\nu\gamma}=\frac{1}{2}G^{\mu\delta}\left(\frac{\partial G_{\delta\nu}}{\partial x^{\gamma}}+
\frac{\partial G_{\delta\gamma}}{\partial x^{\nu}}-\frac{\partial G_{\nu\gamma}}{\partial x^{\delta}}\right)
\end{equation}
is the coefficient of affine connection.
This result indicates that the propagation of light waves in a non-dispersive moving medium is equivalent
to light waves propagating in a curved space-time characterized by the Gordon metric.

However, for a moving dispersive medium, $G^{\mu\nu}=G^{\mu\nu}(x,k)$,
and the refractive index of the medium can be written as $n=n(x,\tilde{\omega})$, where $\tilde{\omega}=u^{\mu}k_{\mu}$
is the local photon frequency in the comoving frame with the medium (hereafter c=1).
In such case, Eq. (A3) is invalid, and can be replaced by
\begin{equation}
\frac{dx^{\mu}}{d\lambda}=K^{\mu}+A^{\mu},
\end{equation}
with
\begin{equation}
A^{\mu}=\frac{1}{2}\frac{\partial G^{\alpha\beta}}{\partial k_\mu}k_{\alpha}k_{\beta}=nn'(u^{\nu}k_{\nu})^2u^{\mu}
\end{equation}
where $dx^{\mu}/d\lambda$ and $K^{\mu}$ are kinetic and canonical momenta for a photon respectively, $A^{\mu}$ is seen
as an effective potential and $n^{'}=\partial n/\partial(u^{\mu}k_{\mu})$ is attributed to the time dispersive effect of the medium
($n^{'}=0$ for the non-dispersive medium). Making use of the Eqs. (A2), (A6) and (A7), and after tediously long
algebraic operation, one can obtain equation of motion for a photon
\begin{equation}
\frac{d^{2}x^{\mu}}{d\lambda^{2}}+\Gamma^{\mu}_{\nu\gamma}\frac{dx^{\nu}}{d\lambda}\frac{dx^{\gamma}}{d\lambda}=f^{\mu}(A)
\end{equation}
and
\begin{equation}
f^{\mu}(A)=\frac{dx^{\nu}}{d\lambda}\frac{\partial A^{\mu}}{\partial x^{\nu}}+G^{\mu\gamma}\frac{dx^{\nu}}{d\lambda}
\left(\frac{\partial G_{\gamma\delta}}{\partial x^{\nu}}-\frac{\partial G_{\nu\delta}}{\partial x^{\gamma}}\right)A^{\delta}
+\frac{1}{2}G^{\mu\gamma}\frac{\partial G_{\delta\nu}}{\partial x^{\gamma}}A^{\nu}A^{\delta},
\end{equation}
where the four vector $f^{\mu}(A)$ is seen as an external field for photons consisting of the effective potential $A^{\mu}$
and its coupling with the tensor $G_{\mu\nu}$. This result displays that a photon is moving under the joint action of the
effective metric $G_{\mu\nu}$ and an external field specified by the effective potential $A^{\mu}$.
The refractive index can be represented as $n^2=1+\chi$.
When the susceptibility of the medium $|\chi|\ll1$, and its derivatives
$(\partial\chi)/(\partial(u^{\mu}k_{\mu}))\ll1$ and $(\partial\chi)/(\partial x^{\mu})\ll1$,
Eq. (A8) can be reduced to the first order in $|\chi|$
\begin{equation}
\frac{d^{2}x^{\mu}}{d\lambda^{2}}+\Gamma^{\mu}_{\nu\gamma}\frac{dx^{\nu}}{d\lambda}\frac{dx^{\gamma}}{d\lambda}
=\frac{dx^{\nu}}{d\lambda}\frac{\partial A^{\mu}}{\partial x^{\nu}}.
\end{equation}
Removing the term $dA^{\mu}/d\lambda$ from the right side of Eq. (A10) to the left side, the equation can be written as
\begin{equation}
\frac{dK^{\mu}}{d\lambda}+\Gamma^{\mu}_{\nu\gamma}\frac{dx^{\nu}}{d\lambda}\frac{dx^{\gamma}}{d\lambda}=0.
\end{equation}
On account of the magnitude of $\Gamma^{\mu}_{\nu\gamma}$ being the first order in $\chi$ as the same order as the effective
potential $A^{\mu}$, one can further to write Eq. (A11) as
\begin{equation}
\frac{dK^{\mu}}{d\lambda}+\Gamma^{\mu}_{\nu\gamma}K^{\nu}\frac{dx^{\gamma}}{d\lambda}=0.
\end{equation}
This results indicates that the canonical momentum $K^{\mu}$ of the photon is parallel-transported along the world lines (light rays) of a photon.

As a summary of the above analysis, light traveling in a dispersive medium is equivalent to that governed under the mutual actions of an
effective metric field and an external field specified by an effective potential attribute to the dispersive effect.
The kinetic momentum for photons, i.e. light rays, is no longer following the null geodesics due to the dispersive effect
of the medium. In the cases of weak susceptibility, to the lowest order in $|\chi|$, the canonical momentum of a photon
is governed by the null geodesics equation
and the Gordon metric $G_{\mu\nu}=g_{\mu\nu}+(1/n^{2}-1)u_{\mu}u_{\nu}$
is still valid as an effective metric in the cases of light propagating in a dispersive medium,
but it is required that an effective electromagnetic potential is added because of the dispersive effect.

\label{lastpage}
\end{document}